
\input harvmac

\overfullrule=0pt


\def\IR{\relax{\rm I\kern-.18em R}}


\def\a{\alpha}

\def\e{\epsilon}

\def\m{\mu}
\def\n{\nu}

\def\l{\lambda}


\noindent hep-th/9205037 \hfill USC-92/HEP-B1
\bigskip

\centerline    {\bf GLOBAL ANALYSIS OF NEW GRAVITATIONAL SINGULARITIES }
 \centerline { {\bf IN STRING AND PARTICLE THEORIES }
               {\footnote {$^*$} {Research supported in part by DOE, under
Grant No. DE-FG03-84ER-40168 }  }
               {\footnote {$^\dagger$} {This paper is dedicated to the
memory of Feza G\"ursey who introduced one of us to the beauty of symmetry and
group theory.  } }       }


\vskip 1.00 true cm

\centerline { I. BARS and K. SFETSOS}

\bigskip

\centerline {Physics Department, University of Southern California}
\centerline {Los Angeles, CA 90089-0484, USA}

\vskip .5 true cm
\centerline{ABSTRACT}
\medskip

We present a global analysis of the geometries that arise in non-compact
current algebra (or gauged WZW) coset models of strings and particles
propagating in curved space-time. The simplest case
is the 2d black hole. In higher dimensions these geometries describe new and
much more complex singularities. For string and particle theories (defined in
the text) we introduce general methods for identifying global coordinates and
give the general exact solution for the geodesics for any gauged WZW model
for any number of dimensions. We then specialize to the 3d geometries
associated with $SO(2,2)/SO(2,1)$ (and also $SO(3,1)/SO(2,1)$) and discuss in
detail the global space, geodesics, curvature singularities and duality
properties of this space. The large-small (or mirror) type duality property is
reformulated as an inversion in group parameter space. The 3d global space has
two topologically distinct sectors, with patches of different sectors related
by duality. The first sector has a singularity surface with the topology of
``pinched double trousers". It can be pictured as the world
sheet of two closed strings that join into a single closed string and then
split into two closed strings, but with a pinch in each leg of the trousers.
The second sector has a singularity surface with the topology of ``double
saddle", pictured as the world sheets of two infinite open strings that come
close but do not touch. We discuss the geodesicaly complete spaces on each
side of these surfaces and interpret the motion of particles
in physical terms. A cosmological interpretation is suggested and comments are
made on possible physical applications.

\bigskip\noindent
PACS: 11.17.+y, 04.20.Jb, 02.40.+m   \hfill               May 1992

\vfill\eject


\newsec{Motivation}

In an attempt to formulate solvable models of strings propagating in curved
spacetime it was discovered that certain classes of non-compact current algebra
models
 \ref\BN{I. Bars and D. Nemeschansky, Nucl. Phys. {\bf B348} (1991) 89.}
can shed light on gravitational singularities such as black holes in
string theory
 \ref\WIT{E. Witten, Phys. Rev. {\bf D44} (1991) 314.} as well as more
interesting singularities in higher dimensions
 \ref\CRE{ M. Crescimanno, Mod. Phys. Lett. {\bf A7} (1992) 489.}
 \ref\BS{I. Bars and K. Sfetsos,
  Mod. Phys. Lett. {\bf A7} (1992) 1091.}
 \ref\BASF{I. Bars and K. Sfetsos, Phys. Lett. {\bf 277B} (1992) 269.}
 \ref\FL{E. S. Fradkin and V. Ya. Linetsky, Phys. Lett. {\bf 277B} (1992) 73.}
 \ref\HORV{P. Horava, Phys. Lett. {\bf 278B} (1992) 101.}
 \ref\GER{D. Gershon, ``Exact Solutions of Four-Dimensional Black Holes
 in String Theory", TAUP-1937-91.}.
All single time coordinate models based on simple non-compact groups are
characterized
 by the $G/H$ cosets \BN\ref\IBCS{I. Bars, ``Curved Space-Time
Strings and Black Holes", in Proc. of {\it XX$^{th}$ Int. Conf. on Diff.
Geometrical Methods in Physics}, Eds. S. Catto and A. Rocha, Vol.2, p.695,
(World Scientific, 1992).} :
 \ $ SO(d-1,2)/SO(d-1,1)$, $\ SO(d,1)/SO(d-1,1)$, $\ SU(n,m)/SU(n)\times
SU(m)$, $\ \ SO(n,2)/SO(n)$, $\ \ SO(2n)^*/SU(n)$, $\ \ Sp(2n)^*/SU(n)$, $\ \
E_6^*/SO(10)$, $\ \ E_7^*/E_6$. Naturally, by taking non-simple (direct
product) groups, including non-Abelian, $U(1)$ or {\IR} factors one can
construct extensions of these models \BN\ provided one demands a total
Virasoro central charge $c=26$ (or $c=15$ with supersymmetry). In addition,
when there are $U(1)$ or {\IR} factors there is a non-trivial way of gauging
a total $U(1)$ or {\IR} that leads to further models. For example, by
combining Witten's 2d black hole with additional $U(1)$'s, black holes, black
strings and black p-branes can be constructed in higher dimensions
 \ref\HH{J. H. Horn and G. T. Horowitz, Nucl. Phys. {\bf B368} (1992) 444.}
 \ref\GS{Giddings and Strominger, Phys. Rev. Lett. {\bf 67} (1991) 2930.}
 \ref\RA{E. Raiten, ``Perturbation of a Stringy Black Hole'', Fermilab-Pub
 91-338-T. }
 \ref\GIN{P. Ginsparg and F. Quevedo, ``Strings on Curved Space-Times:
 Black Holes, Torsion, and Duality'', LA-UR-92-640.}.
In all of these models there are interesting duality properties
 \ref\DUAL{ A. Giveon, Mod. Phys. Lett. {\bf A6} (1991) 2843.
              \semi
 I. Bars,``String Propagation on Black Holes'', USC-91/HEP-B3.\semi
 R. Dijkgraaf, H. Verlinde and E. Verlinde,
Nucl. Phys. {\bf B371} (1992) 269. \semi
 E. Kiritsis, Mod. Phys. Lett. {\bf A6} (1991) 2871.      }
that generalize the kind of duality found in tori and mirror manifolds.
The existence of a {\it discrete} duality and its origins in the general model
 \foot{It must be emphasized that the cosets $G/H$ are not the left or right
cosets in which the subgroup $H$ acts on one side of the group element.
Rather, $H$ is the ``vector" subgroup that acts on both sides $g\rightarrow
\Lambda g\Lambda^{-1}$, where $\Lambda\in H$. Actually, $H$ may be generalized
\BS\ to a deformed subgroup $g\rightarrow \Lambda g \tilde \Lambda^{-1}$ where
$\tilde \Lambda =g_0\Lambda g_0^{-1}$ or
$\tilde \Lambda =g_0(\Lambda^T)^{-1} g_0^{-1}$ and $g_0$ is a discrete element
in {\it complexified} $G$. This deformation is the origin of duality
transformations. Its effects can be reproduced as discrete leaps in group
space $g\rightarrow \hat g$ by the switching of some signs. In this latter
sense the undeformed theory already describes all
the dual sectors. More on this point will be said below.}
 is given in \BS\ , while further duality properties based on Killing vectors
is given in
 \ref\VR{M. Rocek and E. Verlinde, ``Duality, Quotiets, and Currents'',
 ITP-SB-91-53. }
 \ref\GR{A. Giveon and M. Rocek, ``Generalized Duality in Curved String
 Backgrounds'', IASSNS-HEP-91/84.}.

By virtue of being exact conformal models, all of these cosets yield new
explicit solutions of Einstein's equations coupled to matter in $d=dim(G/H)$
dimensions by automatically solving the perturbative conformal invariance
conditions of
 \ref\CAL{C. G. Callan, D. Friedan, E. J. Martinec and M. Perry, Nucl.
Phys. {\bf B262} (1985) 593.}.
Recall the traditional Lagrangian method for constructing the metric that
solves Einstein's equations: start with a gauged WZW model characterized by
G/H
 \ref\WZW{E. Witten, Nucl. Phys. {\bf B223} (1983) 422.
 \semi K. Bardakci, E. Rabinovici and B. Saering, Nucl. Phys. {\bf B301}
(1988) 151.
 \semi K. Gawedzki and A. Kupiainen, Nucl. Phys. {\bf B320} (1989) 625.
 \semi H.J. Schnitzer, Nucl. Phys. {\bf B324} (1989) 412.
  \semi D. Karabali, Q-Han Park, H.J. Schnitzer and Z.Yang, Phys.Lett.
 {\bf 216B} (1989) 307.
 \semi D. Karabali and H.J. Schnitzer, Nucl. Phys. {\bf B329} (1990) 649.}
,
parametrize the group element, choose a unitary gauge that fixes the gauges of
H completely by setting $dimH$ group parameters equal to zero and integrate
out the non-propagating $dimH$ gauge fields.  The remaining group parameters
$q^a(\tau,\sigma),\ a=1,2,\cdots,dim(G/H)$ are interpreted as target space
string coordinates and the Lagrangian looks like a non-linear sigma model that
describes a string propagating in $dim(G/H)$ curved spacetime
 \foot{This procedure which was briefly used in \BN\ to count the number of
time coordinates, was fully explored for $SL(2,\IR)$ in \WIT\ to construct the
2d black hole metric.}.
 The gravitational metric $G_{ab}(q)$ is identified and interpreted by
examining this effective Lagrangian.  Similarly, one can read off a dilaton
field $\Phi(q)$ directly from the WZW lagrangian \BS\BASF . Remarkably, this
metric and dilaton, together with an antisymmetric field $B_{ab}(q)$ which
also emerges in a similar way (sometimes zero), automatically solve the
coupled Einstein's equations for $G_{ab},B_{ab},\Phi$ in dimension
$d=dim(G/H)$. Therefore, in addition to string theory, this approach is
tempting for investigations in General Relativity since it may be viewed as a
new method for generating fascinating classical solutions to Einstein's
equations. Even more enticing is the fact that the quantum spectrum is solved
exactly by labelling the states group theoretically and computing eigenvalues
of Casimir operators.

In string theory, this metric is perturbative from the point of view of
conformal invariance and is valid at large $k$, where $k$ is the central
extension of the current algebra. In the same large $k$ limit, the algebraic
properties of the cosets above indicate \BN\ that one should expect a
$dim(G/H)$ geometry with a {\it single time coordinate}; indeed this is
satisfied in intricate ways as will be explicitly seen below. As we will show
elsewhere
\ref\BAS{I. Bars and K. Sfetsos, in preparation }
the conformally exact metric and dilaton are computed explicitly by algebraic
Hamiltonian methods for any $k$. For the bosonic and heterotic string there
are major corrections. However, for the supersymmetric type-II string based on
any $G/H$, the exact metric and dilaton are the same as the perturbative
metric and dilaton, except for an overall renormalization ($k$ gets replaced
by by $k-g$, where $g$ is the Coxeter number of $G$). Furthermore, for a
particle theory (as opposed to a string theory, see below) the perturbative
and exact expressions are identical for any $k$ with or without supersymmetry.
Therefore, it is meaningful to study the so called ``perturbative" metric for
a variety of cases.

Another problem with the perturbative Lagrangian method is that it generates
the metric in a patch of the manifold. By choosing a somewhat different
unitary gauge one arrives at a metric, in a different coordinate patch, which
may bear no resemblance to the previous one (e.g. compare \CRE\ to \BS\ or
\FL\ in 3d). What is the global space? What are the global coordinates? What
is the behaviour of light rays (or slower moving particles) in the
geodesically complete space? One needs this information in order to interpret
the geometry of spacetime. In two dimensions this problem was solved by
rewriting the metric in terms of the globaly defined Kruskal coordinates \WIT
. However, in higher dimensions, in the absence of Killing vectors, we need
new methods.

For the reasons mentioned above the new singular geometries that arise in
higher dimensions have not been easy to interpret (except for $U(1)$'s
times 2d black hole). It is the purpose of the present paper to do so. Our
treatment will introduce methods that are completely general and apply to all
of the above models, and in fact to any effective metric derived from a gauged
Wess-Zumino-Witten model (any number of time coordinates 0,1,2, etc.).

\newsec{Global space and geodesics in the general model}

Our first observation is that the global coordinates must be $H$-invariant.
This will avoid the problem with gauge fixing. In fact, although not
immediately obvious, there are precisely $dim(G/H)$ independent H-invariants
$Q^a$ that can be constructed from $dimG$ group parameters. To illustrate this
point, consider $SO(d-1,2)/SO(d-1,1)$. Under the Lorentz subgroup
$H=SO(d-1,1)$ the group parameters are classified as a vector $x_\mu$ and an
anti-symmetric tensor $a_{\mu\nu}$ in $d$-dimensions. There are precisely $d$
Lorentz invariants that can be constructed from these. For example for $d=3$
we have $Q^a$=($x^2$, $a^2$, $\epsilon^{\mu\nu\lambda}x_\mu
a_{\lambda\sigma})$, for $d=4$ we have $Q^a$=($x^2$ , $a^2$,
$\epsilon^{\mu\nu\lambda\sigma}a_{\mu\nu} a_{\lambda\sigma}$, $(x^\mu
a_{\mu\nu})^2$), etc.. The invariants $Q^a$ are related to the gauge fixed
group parameters $q^a$ by a change of coordinates $q^a(Q)$, as will be
shown below. This example also illustrates why one may get only a patch
if the metric is written in terms of gauge fixed group parameters: when $x^2$
is time-like it can be Lorentz transformed (or gauge fixed) to $x_\mu
=(x_0,0,0,\cdots)$ but if it is space-like or light-like it cannot be put into
this form. Thus, the metric will look very different and cover different
patches for these two possibilities. However, as an invariant, $x^2$ can take
zero, positive and negative values, thus ``globally" covering all
possibilities. Similar comments apply to all other invariants.

How does one rewrite the metric in terms of H-invariants? We have found an
algebraic and systematic approach which also leads to many other results, such
as the exact conformal metric, dilaton, etc.. Because it involves a host of
other ideas and techniques it will be published as a separate article \BAS\ .
Here we will use a more pedestrian approach which agrees with our systematic
results. We start with the metric which is computed in any patch and rewrite
the group parameters in terms of dot products of {\it gauge fixed} $H$-
representations. We make a change of variables from gauge fixed group
parameters to these dot products ($dim(G/H)$ of them) and then allow the new
coordinates thus identified to take all possible values that invariants can
take. This procedure will provide the needed analytic continuation from the
original patch to the global space. Then the location and nature of the
singularities in the geometry are revealed.

Having global coordinates is not sufficient to get a feeling for the geometry;
one also needs to know the behavior of the geodesics. However, the geodesic
equation seems completely unmanageable in the complicated metrics that emerge.
On the other hand, we have been able to find the general geodesic solution by
the following procedure. We first define a gauged point particle theory, which
is essentially the dimensional reduction of the familiar WZW model (i.e. all
our fields are functions of only $\tau$, rather than $\tau,\sigma$).

\eqn\action{ S(g)={k\over 4\pi}\int d\tau \ Tr\big ({1\over 2}g^{-1}\dot{g}\
g^{-1}\dot{g} - A_-\dot{g}g^{-1} - A_+g^{-1}\dot{g} + A_-g A_+g^{-1}-A_-
A_+\big ) }
where $g(\tau)\in G$ is a group element and $A_{\pm}(\tau)$ are two gauge
potentials in the Lie algebra of $H$. Two gauge potentials are needed for our
purposes. The model is gauge invariant under the transformations
$g'=\Lambda g \Lambda^{-1},\ \ A'_{\pm}=\Lambda (A_{\pm}-\partial_\tau)
\Lambda^{-1}$, where $\Lambda (\tau) \in H$. Consider the equations of motion
\foot{The third and fourth equations are derived from a more complicated
equation after using the first one and projecting along $H$ or $G/H$ in the Lie
algebra.}

\eqn\eqs{ (g^{-1}D_-g)_H =0= (D_+gg^{-1})_H, \quad D_+(g^{-1}D_-g)=0, \quad
\dot{A}_--\dot{A}_+={1\over 2}[A_++A_-,A_--A_+] . }
where $D_\pm g=\dot{g}-[A_\pm , g]$, and the subscript $H$ means a projection
to the Lie algebra of $H$. These equations may be considered as geodesics in
an enlarged space ($dimG+2dimH$). One avenue for solving these equations is to
choose a unitary gauge ($dimH$ conditions), solve for the two gauge potentials
and substitute in the remaining equations. The remaining unsolved $dim(G/H)$
equations are in fact the geodesic equations. That is, these are the equations
of motion that follow from the $dim(G/H)$ non-linear sigma model (equivalent
to \action ) that defines the line element for the metric, $ S=(k/\pi)\int
d\tau \ G_{ab}(q)\ \dot{q}^a\dot{q}^b+\cdots$, and they coincide with the
usual geodesic equation for that metric,
$\ddot{q}^a+\Gamma_{bc}^a\dot{q}^b\dot{q}^c=0$. We may rewrite these geodesic
equations in terms of the global coordinates $Q^a(q)$ described above. It
seems hopeless to find solutions for them (see e.g. (6.1)). However, another
avenue for solving \eqs\ is to choose an axial gauge $A_+=0$. In this gauge
there is a leftover global $H$-invariance giving all expressions a $H$-
covariant form. The last equation yields $A_-=\alpha$ where $\alpha$ is a
constant matrix in the Lie algebra of $H$. The first and third equations yield
a first integral of the form $g^{-1}D_-g=p$, where $p$ is a constant matrix in
the Lie algebra of $G/H$. This equation can be rearranged to the form
$\dot{g}=g(p-\alpha)+\alpha g$, and then solved by

\eqn\sol{ g=e^{\alpha\tau}g_0e^{(p-\alpha)\tau}, }
where $g_0$ is a constant group element that characterizes the initial
conditions. Finally, replacing this form into the remaining second equation in
\eqs\ yields a constraint among the constants of integration

\eqn\const{  \big (g_0(p-\alpha)g_0^{-1}\big )_H+\alpha =0 .}
Let us count the constants of integration. We start with
$dim(G)+dimH+dim(G/H)$ parameters in $g_0,\alpha,p$. The constraints and
the leftover global $H$-invariance remove $2dimH$ of them. Therefore, the
truly independent and physical ones are $2dim(G/H)$ in number, which is
precisely the number of initial positions $Q^a(0)$ and velocities
$\dot{Q}^a(0)$ needed for the general geodesic in $dim(G/H)$ dimensions.
Therefore \sol\const\ contain the general geodesic solution. What remains is
the purely group theoretical exercise of projecting from this solution in
group space $G$ to the space of $H$-invariants and relating them to the
coordinates $Q^a(\tau)$. These will then give the general geodesic solution in
the {\it global} space! It was very important that we reformulated the
manifold in terms of $H$-invariants because, by virtue of being gauge
independent, the solution obtained for the invariants in the axial gauge is
indeed the solution in any gauge, and in particular in all the patches of the
unitary gauge where the question was first asked.

As already mentioned, in a unitary gauge the Lagrangian \action\ is rewritten
in terms of the line element $ds^2/d\tau^2=\dot{q}^a\dot{q}^b G_{ab}(q)$.
Therefore, if we substitute the covariant solution \sol\const\ in the gauge
invariant Lagrangian \action\ we find the value of $ds^2/d\tau^2$ for the
geodesic solution. This gives

\eqn\dsds{ {ds^2\over d\tau^2}= {1\over 2} Tr(p^2) . }
Now by choosing the constant matrix $p$ we have control on whether the
geodesic is light-like, time-like or space-like. This feature will allow us to
examine below the behavior of light rays in the curved geometries that emerge
by taking $Tr(p^2)=0$.

The above solution was for the self-contained particle theory of \action . The
string theory has a similar fully general solution as given in \BS .
Therefore, we are also equipped to study  the string geodesics in these
geometries. In fact, by applying techniques similar to those displayed below
we can find out the general string motion in curved spaces containing
singularities (such as black holes) provided they are generated through a WZW
model. The solution exhibited in \sol\const\ simply corresponds to the motion
of a string collapsed to a point. In this paper we will not elaborate on
the more general string geodesics and invite the interested reader to
synthesize the solutions of \BS\ with the methods of this paper. In this way
one can study, for example, the free fall of a string into a black hole.

\newsec{Global space for 3d $SO(2,2)/SO(2,1)$ and $SO(3,1)/SO(2,1)$}

In this section we will apply the general ideas to the specific cases
$SO(2,2)/SO(2,1)$ and $SO(3,1)/SO(2,1)$ to find the global 3d geometry. To
begin we will adopt the geometrical analysis of the $SO(2,2)/SO(2,1)$ string
theory given in \BS\ which applies unchanged to the point particle theory
defined by \action\ . The $SO(3,1)/SO(2,1)$ case will be discussed by
analytic continuation after the global space for $SO(2,2)/SO(2,1)$ is
understood. The $SO(2,2)$ group element was written in terms of 6
parameters that are classified as two Lorentz 3d-vectors ($X_\mu,Y_\mu$) under
the subgroup $SO(2,1)$. The group element takes the form g=h(Y)t(X), where

\eqn\htt { \eqalign {
 &h=\left ( \matrix {1 & 0 \cr 0 & h_\mu^{\ \nu} \cr } \right ) ,
 \qquad \qquad t=\left (\matrix {b  & -bX^\nu \cr
    bX_\mu  & (\eta_\mu^{\ \nu} + ab X_\mu X^\nu) \cr } \right ),  \cr
 &h_{\mu\nu}=\epsilon' \sqrt{1-Y^2}\eta_{\mu\nu}+{Y_\mu Y_\nu\over 1+\epsilon'
\sqrt{1-Y^2} } + \epsilon_{\mu\nu\lambda}Y^\lambda, \cr
 & b={\epsilon\over \sqrt{1+X^2} }, \qquad a=(1-b^{-1})/X^2, \qquad
\epsilon=\pm,\ \epsilon'=\pm, \qquad\eta_{\mu\nu}=diag(1,-1,-1) . } }
Indices are raised or lowered with the Minkowski metric $\eta_{\mu\nu}$. The
gauge transformations are Lorentz transformations. The unitary gauge that
describes a patch was fixed by taking $X_\mu=tanh(2r)\ (0,0,1)$ and
$Y_\mu=sinh(2t)\ (0,cos\theta ,-sin\theta )$, in which both invariants
($X^2,Y^2$) are negative (space-like). After solving for $A_\pm$ and
substituting in the action, and taking care of the correct measure in the path
integral \BS\BASF, the effective Lagrangian takes the form
$L=(k/\pi)(ds/d\tau)^2 - \Phi $ where the line element $ds^2$ and the dilaton
$\Phi$ are given in terms of the coordinates $q^a=(t,r,\theta)$

\eqn\metric {\eqalign {
 & ds^2=dr^2+\lambda^2(r,\epsilon)\big[d\theta +\kappa (t,\epsilon')\
tan\theta\ dt \big]^2-\lambda^{-2}(r,\epsilon)\ cos^{-2}\theta\ dt^2 \cr
 & \lambda^2(r,\epsilon)={cosh(2r)-\epsilon\over cosh(2r)+\epsilon},\qquad
 \kappa(t,\epsilon')={sinh(2t)\over cosh(2t)-\epsilon'},\qquad \epsilon=\pm,\
 \epsilon'=\pm \cr
 & \Phi (r,t,\theta)=ln \big [sinh^2(2r)\ cos^2(\theta)\ (cosh(2t)-
\epsilon')\big ] + \ constant .            }  }
For the particle theory this gives the exact metric and dilaton for any $k$.
Also, as shown in \BAS\ , this is the exact metric and dilaton for the
supersymmetric $SO(2,2)/SO(2,1)$ {\it type-II superstring} for any $k$,
except for an overall quantum renormalization that replaces $k$ by $(k-2)$ in
the effective Lagrangian. However, for the purely bosonic or the heterotic
string theory this is the large $k$ conformally perturbative metric and
dilaton, while the conformally exact expressions for any $k$ given in \BAS\
differ from the above.

The various signs $\epsilon ,\epsilon'$ correspond to
patches related by duality transformations that will be discussed further
below. For every $\epsilon,\epsilon'$, there are additional sets of analytic
continuations into other patches that correspond to other configurations of
unitary gauges in different regions of $X^2,Y^2$. The 16 patches so obtained
form the global space as shown below.

The goal now is to rewrite the metric in terms of an appropriate
combination of dot products $(X^2,Y^2,X\cdot Y)$ in such a way that a single
expression for the metric is valid in all 16 patches. This can be done easily
in one patch, and then the expression so obtained is allowed to take values
for the entire range of the invariants. This gives the global metric in the
global space. We find it convenient to define the following invariants
suggested by the form of the group element in \htt

\eqn\inv{ b=\epsilon (1+X^2)^{-1/2}, \qquad v=\epsilon'(1-Y^2)^{1\over 2}+1,
\qquad u=(v-2) (X\cdot Y)^2/X^2Y^2 . }
In the unitary gauge for the patch above these reduce to

 \eqn\red{ b=\epsilon\ cosh(2r), \qquad v=\epsilon' cosh(2t)+1, \qquad
u=sin^2\theta\ (\epsilon' cosh(2t)-1). }
Using this last equation as a change of coordinates $Q^a(q)$ we rewrite the
metric and dilaton  \metric\ in terms of $Q^a=(v,u,b)$

\eqn\nmetric{\eqalign {
 & ds^2={db^2\over 4(b^2-1)}+{b-1\over b+1}{du^2\over 4u(v-u-2)}-
{b+1\over b-1}{dv^2\over 4v(v-u-2)}, \cr
 & \Phi = ln[(b^2-1)(v-u-2)]+\Phi_0 . } }
The non-zero components of the Ricci tensor are given by

\eqn\ricci{ R_{bb}=-{2\over (b^2-1)^2},\qquad R_{vv}={2bv-(b+1)^2\over v(b-
1)^2(v-u-2)^2},\qquad R_{uu}={-2bu+(b-1)^2\over u(b+1)^2(v-u-2)^2}, }
while the scalar curvature for this metric is

\eqn\scalar{ R=8{3+b^2-v(b+1)-u(b-1)\over (b^2-1)(v-u-2)}, }
revealing the location of gravitational singularities in the global space. The
discussion of the singularity will be postponed to section 4.

In the regions of the global space for which $|b|\gg 1$ and $|v-u|\gg 2$ the
curvature vanishes, indicating that the metric is flat. For such asymptotic
regions, we may exhibit the flat metric by the reparametrization

\eqn\flat{ \eqalign {
 & b\sim {\epsilon} e^{2z_1}, \qquad v\sim \epsilon'e^{2z_0}
sinh^2z_2, \qquad u\sim \epsilon'e^{2z_0}cosh^2z_2, \cr
 & ds^2\sim dz_1^2+dz_2^2-dz_0^2 } }
which is valid for large values of $z_0,z_1$. Since the {\it time coordinate}
 $z_0$
must get large to reach the asymptotically flat region, this metric {\it
must have a cosmological
interpretation}. This will be discussed in the last section of the
paper.

We now determine the globally allowed ranges of $Q^a=(v,u,b)$ by analytic
continuation away from the patch \red\ . First, the correct parametrization of
the $SO(2,2)$ group element requires that ($X^2>-1,\ Y^2<1$) be within the
ranges which insure that the square roots in \htt\inv\ are real. Furthermore,
the existence of the dual patches allows $Q^a=(v,u,b)$ to take both positive
and negative values. This translates to $b,v$ taking values on the entire real
line. The only remaining task is the determination of all allowed values for
the Lorentz dot products  $(X\cdot Y)^2/X^2Y^2$ when ($X^2,Y^2$) are in their
allowed ranges. It is easy to see that this provides restrictions on the
combined ranges of $(v,u,b)$ as follows:

\eqn\ranges{ [(b^2>1)\ {\rm and}\  (uv>0)],\qquad or\qquad
  [(b^2<1)\ {\rm and}\  (uv<0) ,\  {\rm excluding}\ 0<v<u+2<2 .] }
This is then the global space for the $SO(2,2)/SO(2,1)$ theory! In Figs.1a,b,c
the 16 patches of this space are enumerated as (I), (IIa-IIg), (IIIb), (I'),
(II'b,II'c,II'f,II'g), (III'a,III'b). The patch we started from, with
$\epsilon=+,\ \epsilon'=+$, is denoted by (I) in Fig.1a, as can be verified
from the values of $Q^a=(v,u,b)$ generated by \red .  The planes that slice-up
the space correspond to the values of $Q^a=(v,u,b)$ at which there is a change
of sign for the factors $[(b^2-1), (v-u-2), v, u]$ which appear in the metric
\nmetric . These crucial sign switches determine the signature of the metric
in the various regions of the global space. The signatures in each patch is
given in Fig.1 in the form $(-++)$, $(++-)$, $(+-+)$, which shows the signs of
the factors in front of $(dv^2,du^2,db^2)$ in the line element \nmetric . It
is seen that for any of the $SO(2,2)/SO(2,1)$ patches there is always {\it
only one time coordinate} and two space coordinates, although the role of time
switches between $(v,u,b)$ in the various regions. The patches are grouped
together in six regions I,I',II,II',III,III' as in the parentheses above. As
explained in section 5, each one of the six groups of patches in parentheses
is geodesically complete and geodesically disconnected from the others.

For completeness, it is instructive to see how the 16 patches that make up the
global $SO(2,2)/SO(2,1)$ space \ranges\ are parametrized in various unitary
gauges. The following table provides this information.

$$\vbox   {    \tabskip=0pt \offinterlineskip
\halign to \hsize  { \strut # & \vrule# & \tabskip= 1em plus 2 em
& #\hfil &\vrule#  & #\hfil &\vrule# \tabskip =0pt \cr \noalign {\hrule}
&& $(patch)_{\epsilon,\epsilon'}$ && $X,Y,(u,v,b)$ & \cr
\noalign{\hrule}
\noalign {\smallskip}
&&$(I)_{++},(I')_{+-} $&&$ X_\mu=tanh(2r)\ (0,0,1), \quad
Y_\mu=sinh(2t)\ (0,cos\theta ,-sin\theta ) $ & \cr
&&$ (II'g)_{-+}, (IIg)_{--} $&&$ u=(v-2) sin^2\theta ,\quad
v=\epsilon'cosh(2t)+1 ,\quad b=\epsilon cosh(2r) $ & \cr\noalign{\hrule}
&&$(IIb)_{++},(II'b)_{+-} $&&$ X_\mu=tanh(2r)(0,0,1), \quad
Y_\mu=sinh(2t)(sinh\psi ,0,cosh\psi) $ & \cr
&&$ (III'b)_{-+}, (IIIb)_{--} $&&$ u=(v-2) cosh^2\psi ,\quad
v=\epsilon'cosh(2t)+1 ,\quad b=\epsilon cosh(2r) $ & \cr\noalign{\hrule}
&&$(IIa)_{+\pm},(III'a)_{-\pm} $&&$ X_\mu=tanh(2r)(0,0,1), \quad
 Y_\mu=sin(2\phi)(cosh(t') ,0,sinh(t')) $ & \cr
&&$  $&&$ u=(v-2) sinh^2t',\quad v=\epsilon'cos(2\phi)+1 ,
 \quad b=\epsilon cosh(2r) $ & \cr\noalign{\hrule}
&&$(IIcf)_{\pm +},(II'cf)_{\pm -} $&&$ X_\mu=tan(2t'')(1,0,0) ,\quad
 Y_\mu=sinh(2r')(sinh\psi ,0,cosh\psi) $ & \cr
&&$  $&&$ u=(v-2) sinh^2\psi ,\quad v=\epsilon'cos(2r')+1 ,
\quad b= \epsilon cos(2t'') $ & \cr\noalign{\hrule}
&&$(IIde)_{\pm\pm} $&&$ X_\mu=tan(2t'')(1,0,0) , \quad
Y_\mu=sin(2\phi)(cosh(r'') ,0,sinh(r'')) $ & \cr
&&$   $&&$ u=(v-2) sinh^2\psi ,\quad v=\epsilon'cos(2\phi)+1 ,\quad
 b= \epsilon cos(2t'') $ & \cr\noalign{\hrule}
}}$$

Fig.1 shows additional regions with signatures that correspond to
zero time coordinates $(+++)$, two time coordinates $(--+), (-+-), (+--)$ and
three time coordinates $(---)$. These cannot be reached from an $SO(2,2)$ group
element. However, by making an analytic continuation to the spaces
$SO(3,1)/SO(3)$ or $SO(3,1)/SO(2,1)$ or $SO(4)/SO(3)$ these geometries can be
described by the same global metric in \nmetric . The corresponding patches
are indicated on Fig.1a,b,c. Note that the two purely
(space,space,space)=$SO(3,1)/SO(3)$ regions are non-compact since $b>1$ or
$b<-1$. This is the region that would be reached from $SO(2,2)/SO(2,1)$ by the
usual Minkowski-Euclidean continuation of Lorentz vectors and tensors which
imply $X_\mu=(X_0,X_1,X_2)\rightarrow (iX_0,X_1,X_2)$ and
$Y_\mu=(Y_0,Y_1,Y_2)\rightarrow (Y_0,iY_1,iY_2)$ (recall
$A^{\mu\nu}=\epsilon^{\mu\nu\lambda}Y_\lambda$). This fits well with the
current algebra approach to string theory: the conformal field theory that
describes strings have the same Virasoro central charge for both of these
cases $c=3k^2/(k-1)(k-2)$, and for $c=26$ a positive value of
$k=(39\pm\sqrt{325})/23$ is needed. The positive sign of $k$ is, of course,
crucial in determining the signature of the metric \BN\ since $k$ is an
overall factor that multiplies the metric in the Lagrangian
 \foot{The computation of the conformally exact metric \BAS\ introduces a
renormalization in the overall factor $k\rightarrow (k-2)$. Therefore, to
maintain the correct signature we must require $k-2>0$ which, in turn, demands
that we take the positive square root $k=(39+\sqrt{325})/23$. In the
supersymmetric theory the central charge is $c=9k/2(k-2)$ which produces
$c=15$ for $k=20/7$. }.

The $SO(4)/SO(3)$ and $SO(3,1)/SO(2,1)$ regions with signatures (time, time,
time) and (time, time, space), etc., do not seem to make physical sense
because of the appearance of more than one time coordinate. However, by
changing the sign of $k$ one can convert these to spaces with signatures
(space, space, space)=$SO(4)/SO(3)$ or (time, space, space) =$SO(3,1)/SO(2,1)$
which do make sense physically, and which are the Euclidean-Minkowski
continuations of each other. This is again in accordance with the counting of
time coordinates in the current algebra approach to string theory \BN\IBCS .
The Virasoro central charge now takes the form $c=3K^2/(K+1)(K+2)$ for a
positive $K$ (i.e. $k=-K$), and has an upper bound of $3$. To construct a
critical conformal field theory these curved spaces have to be combined with
additional spaces in order to reach the critical value of $c=26$. So, the
global geometry for the critical string theory depends also on the additional
spaces. Thus, the $SO(3,1)/SO(2,1)$ or $SO(4)/SO(3)$ based models cannot
correspond to a string theory in purely 3 dimensions under any circumstances
and have no relation to the $c=26$ 3d-string theory based just on
$SO(2,2)/SO(2,1)$ or its Euclidean continuation $SO(3,1)/SO(3)$.

For comparison of these results to the 2 dimensional case based on
$SO(2,1)/SO(1,1)$ or the Euclidean continuation $SO(1,2)/SO(2)$, as well as to
$SO(3)/SO(2)$, we have written an Appendix.

\newsec {Duality in 3d}

In this section we comment on the duality properties of the $SO(2,2)/SO(2,1)$
manifold.  The general group theoretical origins of duality transformations
was explained in \BS . For the 3d model the duality transformations are
generated by switching the signs $\epsilon , \epsilon'$. This is equivalent to
considering related dual models where the gauge group is a deformed subgroup
(see footnote 1), as in the vector versus axial U(1) in 2d. In any case, the
global space of any of these models already contains all the dual regions, and
provided it is fully identified as in the previous section, it is sufficient
to consider only the undeformed vector subgroup. For comparison, the duality
properties in 2d are discussed in Appendix A, including its reformulation as
an inversion in group parameter space.

The duality transformations in the 3d manifold which are
generated by $\epsilon$ or $\epsilon'$ may be rewritten in the form

\eqn\duality{ (v',u',b')=(v,u,-b) , \qquad (v'',u'',b'')=(2-v,\ uv/(2-v),\ b).
}
The first duality transformation in \duality\
generated by $\epsilon$ flips the following pairs of primed and unprimed
patches into each other: (I,II'g), (IIa,III'a), (IIb,III'b), (IIg,I'),
(IIIb,II'b) while also flipping the pairs of unprimed patches (IIcd,IIfe)  and
primed patches (II'c,II'f). Similarly, the second duality transformation in
\duality\ transforms the following pairs of patches into each other (I,I'),
(IIb,II'b), (IIc,II'c), (IIg,II'g), (IIf,II'f), (IIIb,III'b) while  sending
the following patches into themselves (IIa), (IIde), (III'a), as indicated on
Fig.1. These generalize the duality properties of the 2d $SO(2,1)/SO(1,1)$
black hole space. As seen from the parametrization of the
$SO(2,2)$ group  element in \htt\ each duality transformation generated by
$(\epsilon,\epsilon')$ makes a discrete leap in $SO(2,2)$ group space. It is
interesting to elaborate on these properties by making a change of group
parameters

\eqn\change{ X_\mu = {2x_\mu\over x^2-1},\qquad Y_\mu = {2y_\mu\over y^2+1} . }
As will be seen in the next section the new parameters $(x_\mu,y_\mu)$ are
natural for expressing the group element in the spinor representation of
$SO(2,2)$ just as the old variables were natural for expressing the group
element in the vector representation of $SO(2,2)$ as in \htt . The allowed
regions $(X^2>-1,\ Y^2<1)$ are reproduced by letting $x^2,y^2$ take values
anywhere on the real line, that is, $-\infty<(x^2,y^2)<\infty$. The global
variables $Q^a=(v,u,b)$ become

\eqn\glo{ b={1-x^2\over 1+x^2},\qquad v={2\over 1+y^2}, \qquad
 u={-2(x\cdot y)^2\over x^2(1+y^2)} .}
{}From these expressions we figure out that the duality transformation
generated
by \duality\ simply corresponds to inversions in $(x_\mu,y_\mu)$ space

\eqn\smalarge{(x'_\mu,y'_\mu)=(-{x_\mu\over x^2},y_\mu),\qquad
                  (x''_\mu,y''_\mu)=(x_\mu, {y_\mu\over y^2}) . }
Note that the $(X_\mu ,Y_\mu)$ given in \change\ remain {\it invariant} under
the duality transformations in \smalarge\  while the group element in \htt\
makes the duality leap just as required by the sign switches of $\epsilon
,\epsilon'$. Again, it is striking how much, these inversion or reflection
forms of duality that we have exhibited, resemble the $R\rightarrow 1/R$
duality of tori or the duality of mirror manifolds (in this connection see
also footnote 1).

\newsec{Pinched Double Trousers and Double Saddle Singularities in 3d.}

Let us now analyze the properties of the curvature singularity. From \nmetric\
and \scalar\ it is seen that the curvature scalar, the dilaton and metric blow
up when

\eqn\blow{ S \equiv (b^2-1)(v-u-2)=0 .}
Evidently, the singularity resides on the planes $b=1,\ b=-1,\ v=u+2$ that
can be imagined easily from Fig.1a,b,c when one thinks in three dimensions
(equivalently, one has $x^2=0,\ x^2=\infty,\ (x.y)^2=x^2y^2$ respectively).
However, we need to do a little more to unravel some multiple sheeted regions
caused by the coordinate singularities at $u=0=v$. It is beneficial to
eliminate the coordinate singularities in order to open up regions that are
folded into ``double sheets" as in the 2d case. The required reparametrization
involves taking square roots $\sqrt{|u|},\sqrt{|v|}$, but since $u,v$ can have
all signs this needs to be done carefully in various regions so that the
global property of the coordinates are preserved. With this in mind let us
examine the global variables in the form \glo\ and construct the combinations
$(b+1)v=4(1+x^2)^{-1}(1+y^2)^{-1}$ and $(b-1)u=4(x\cdot y)^2(1+x^2)^{-
1}(1+y^2)^{-1}$. From this one concludes that $(b+1)v, \ (b-1)u$ have the same
sign. Examining Fig.1a,b,c one sees that in the unprimed regions I,II,III they
are positive and in the primed regions I',II',III' they are negative. We will
call them the positive and negative regions respectively. Recall that every
primed region is dual to some patch of the unprimed one. As we shall further
see in section 5 the primed and unprimed regions are geodesically isolated
from each other. These observations allow us to define a new set of global
coordinates $(\lambda_+,\sigma_+)$ or $(\lambda_-,\sigma_-)$ separately in the
positive or negative regions respectively

\eqn\newco{ \lambda^2_\pm =\pm v(b+1)=4|(1+x^2)(1+y^2)|^{-1}, \quad
\sigma^2_\pm =\pm u(b-1)=(x\cdot y)^2\lambda^2_\pm,
      \quad  -\infty <\lambda_\pm , \sigma_\pm < \infty .}
In terms of the new coordinates the metric and curvature scalar take the
following forms in the positive and negative regions

\eqn\nnmetric{\eqalign {
 & ds^2= {1\over 2S}\big [\pm db^2+2(1+b)\ d\lambda^2_\pm-2(b-1)\ d\sigma^2_\pm
 + 2\ db (\sigma_\pm d\sigma_\pm - \lambda_\pm d\lambda_\pm) \big ] \cr
 & R=-{8\over S} \big [\lambda^2_\pm +\sigma^2_\pm \pm (b^2+3)\big ] \cr
 & S=\pm 2(b^2-1)-(b-1)\lambda^2_\pm + (b+1)\sigma^2_\pm              }  }
where the singularity factor $S$ is the same as in \blow\ up to a $\mp$ sign.
The disadvantage of this coordinate system is that the metric is not diagonal,
but it has other advantages from the point of view of the geodesics and the
overall intuitive view of the singularity. It is also possible to define
another set of coordinates $(\rho_\pm , \omega_\pm )$ that have a diagonal
metric in the positive and negative regions

\eqn\another{  \rho^2_\pm =\pm v\  sign(b+1) ,\qquad
           \omega^2_\pm =\pm u\ sign(b-1) .}
The metric and curvature now take the form

\eqn\nnnmetric{\eqalign {
 & ds^2={db^2\over 4(b^2-1)} + {1\over S}\ \big [(b+1)|b+1|\ d\rho^2_\pm
                        - (b-1)|b-1|\ d\omega^2_\pm \big ]  \cr
 & R=-{8\over S} [|b+1|\rho^2_\pm + |b-1|\omega^2_\pm \pm (b^2+3)] \cr
 & S=\pm 2(b^2-1) - (b-1)|b+1|\rho^2_\pm + (b+1)|b-1|\omega^2_\pm .       }  }
In Figs.3a,b and Figs.4a,b we show the surface formed by the singularity
$S=0$ in the coordinate systems defined above. These pictures were generated
using Mathematica 2.0. The clearest interpretation is obtained in the
$(\lambda_\pm, \sigma_\pm)$ parametrization.

Let us first discuss the positive region with the $(\lambda_+,\sigma_+,b)$
coordinates. The topology of the surface in Fig.3a is, with some imagination,
that of two propagating closed strings, joining into one closed string, and
then splitting again into two closed strings. The initial and final closed
strings shrink down to a single point just before joining and just
after splitting. We call the singularity surface formed by this system
of strings the {\it pinched double trousers}  singularity. In Fig.4a the
surface, which is parametrized by $(\rho_+,\omega_+,b)$, is deformed into more
regular shapes but retains the same topology of the pinched double trousers.
There is also the three dimensional picture that can be imagined with the aid
of Fig.1a,b,c in which the positive region is ``folded up" and deformed into
the regular 3 dimensional blocks labelled by the various unprimed regions
I,II,III. From Fig.3a one can intuitively see that the many patches of Fig.1
or Fig.4a have formed some apparently connected and disconnected regions which
was not easy to deduce from Figs.1,4. Indeed, the overall feeling conveyed by
Fig.3a about the division of the 3d positive space into a connected region
(outside of the trousers) and disconnected regions (inside of {\it each}
pinched leg) is the correct feeling and it will be confirmed by examining the
geodesics. The inside of the body of the trousers is not reachable by any
geodesic. In fact, this was the $(---)$ region of Fig.1b.

At $b=\pm 1$ the mapping between Fig.3a and Fig.4a is tricky. Since
$\lambda_\pm=\rho_\pm |b+1|^{1\over 2}$ and $\sigma_\pm =\omega_\pm |b-
1|^{1\over 2}$, we see that the singularity which consists of a single line
along either the $\lambda_+$ or the $\sigma_+$ axis in Fig.3a has expanded
into the caps of the cylinder and the bottoms or caps of the hyperboloids in
Fig.4a. The rest of the finite $(\lambda_+,\sigma_+)$ planes at $b=\pm 1$ are
mapped to either $\rho_+=\pm\infty$ or $\omega_+=\pm\infty$. The remainder of
the finite $(\rho_+,\omega_+)$ planes at $b=\pm 1$ in Fig.4a,  although they
are not part of the surface of the trousers, are also sqeezed on top of the
singularity lines at $\lambda_+=0$ or $\sigma_+=0$ in Fig.4a. Thefeore one
wonders about the properties of these parts of the $b=\pm 1 $ planes and in
particular about the singularity. For example what happens if a particle
attempts to cross from $|b|<1$ to $|b|>1$ in these regions of Fig.4a? Our
analysis of geodesics in the next section will address these and other issues.
The answer is that geodesics cannot penetrate from $|b|>1$ to $|b|<1$ when
both $(\rho_\pm,\omega_\pm)$ are finite, but they can do it by moving through
$(\rho_\pm,\pm\infty,1)$ or $(\pm\infty,\omega_\pm,-1)$.  This means that
the rest of the $(\rho_\pm,\omega_\pm)$ planes at $b=\pm 1$ (which are not
shaded in Fig.4a,b) are in fact part of
the singularity except at infinity. On the other hand, in the
$(\lambda_\pm,\sigma_\pm,b)$ coordinates (Fig.3a,b)
the only singularities at $b=\pm 1$
are only along the axes at $\sigma_\pm=0$ (for $b=1$)
or $\lambda_\pm =0$ (for
$b=-1$).

The singularity surface in the negative region of Fig.3b or 4b may be called
the {\it double saddle} singularity. It has the topology of the world
surface of two infinitely long propagating open strings which come close but
do not interact. The upper side of the saddle at $b>1$ corresponds to region
I' of Fig.1a while the lower side of the inverted saddle at $b<-1$ corresponds
to region III'ab of Fig.1c. The space in between the
saddles are the various parts of region II' of Fig.1a,b,c : II'b for $b>1$,
II'g for $b<-1$ and II'ef for $-1<b<1$. As one may deduce intuitively from
these pictures I',II',III' are geodesicaly disconnected, while the various
parts within each region are geodesically connected. This is confirmed by the
geodesic solution.

As explained in section 3, the global space for $SO(3,1)/SO(3)$ is obtained
from the $SO(2,2)$ notation by the Euclidean continuation of
vectors and tensors. This amounts to $x^2\rightarrow -\vec x^2,\ y^2\rightarrow
+\vec y^2,\ x\cdot y\rightarrow i\vec x\cdot\vec y$. Where $\vec x, \vec y$ are
3d Euclidean vectors. Then, $b=(1+\vec x^2)/(1-\vec x^2),\ v=2/(1+\vec y^2)$
and $u=(v-2)(\vec x\cdot\vec y)^2/\vec x^2 \vec y^2$. The regions spanned by
these invariants (with $\vec x^2,\vec y^2\ge 0$) are the triangular regions
with $b>1$ and $b<-1$, as indicated in Fig.1a,c. the $\epsilon$ duality flips
these two regions while the $\epsilon'$ duality sends them to themselves. In
the $(\lambda_\pm,\sigma_\pm,b)$ notation the Euclidean continuation amounts
to $\sigma_\pm\rightarrow i\sigma_\pm$. Similarly in the
$(\rho_\pm,\omega_\pm,b)$ notation Euclidean continuation is equivalent to
$\omega_\pm\rightarrow i\omega_\pm$. The $\pm$ regions now correspond to
$sign(b)=\pm$. Therefore, the expressions for the metric, dilaton, etc. are
obtained from the foregoing $SO(2,2)/SO(2,1)$ expressions
\nnmetric\nnnmetric\ by performing this substitution. The singularity surface
has its simplest shape in the $(\rho_\pm,\omega_\pm,b)$ coordinates. It
consists of two half-infinite cylinders of radius $\sqrt{2}$, axis $b$ and
caps at $b=\pm 1$. The $SO(3,1)/SO(3)$ global space is the inside of these
half cylinders.

Through a similar analysis it is straightforward to discuss the
$SO(3,1)/SO(2,1)$ space. Actually we can obtain all the relevant expressions
by making an analytic continuation from $SO(2,2)/SO(2,1)$ through the
substitutions $x_\mu=(x_0,x_1,x_2)\rightarrow (x'_2,ix'_1,x'_0)$ and  $y_\mu
=(y_0,y_1,y_2)\rightarrow (iy'_2,y'_1,iy'_0)$. The net effect on the invariants
is to replace them by
$(x^2,y^2,x\cdot y)\rightarrow (-x'^2,y'^2,-ix'\cdot y')$.
Therefore, although the metric, dilaton, etc. continue to have the same form
as \nmetric\ , the range covered by the coordinates $Q^a=(v,u,b)$ is different
as indicated on Fig.1a,b,c. There are again positive and negative regions
which are disconnected and related by duality. As seen from \newco\ and
\another\ the metric, dialaton, the singularity surface, etc. are obtained
from \nnmetric\ or \nnnmetric\ by replacing $\sigma_\pm\rightarrow -
i\sigma_\pm$ or $\omega_\pm\rightarrow -i\omega_\pm$. However, as already
pointed out in the previous section, to construct a critical string theory one
needs to combine this space with additional spaces in order to obtain $c=26$
(for point particles this requirement can be relaxed), and therefore it may be
necessary to take into account the global properties of the total space. For
this reason we will refrain from giving further details about this case in
this paper. The $SO(4)/SO(3)$ space is similarly discussed by the Euclidean
continuation of $SO(3,1)/SO(2,1)$.

\newsec{ Geodesics and global properties.}

The geodesic equations that follow from the diagonal line element \nmetric\
are

\eqn\geod{\eqalign {
 & \ddot{b}-{b\dot{b}^2\over b^2-1}-{b-1\over b+1}{\dot{u}^2\over u(v-u-2)}-
{b+1\over b-1}{\dot{v}^2\over v(v-u-2)}=0, \cr
 & \ddot{u}+({1\over v-u-2}-{1\over u}){\dot{u}^2\over 2}-{\dot{u}\dot{v}\over
v-u-2} + {2\dot{u}\dot{b}\over b^2-1} + {(b+1)^2\over (b-1)^2}
{u\dot{v}^2\over 2v(v-u-2)} = 0, \cr
 & \ddot{v}-({1\over v-u-2}+{1\over v}){\dot{v}^2\over 2}+{\dot{u}\dot{v}\over
v-u-2} - {2\dot{v}\dot{b}\over b^2-1} - {(b-1)^2\over (b+1)^2}
{v\dot{u}^2\over 2u(v-u-2)} = 0, \cr
} }
It seems impossible to find a general solution. We might look for special
solutions in which one of the variables is held fixed (e.g. $b$=constant). Note
that the correct equation for this case is to be obtained from the geodesic
equations and not by first specializing the line element (e.g. $db$=0).  There
is a difference between first varying the action and then setting a variable
to a constant (correct procedure) versus first setting a variable to a
constant and then varying the action in the remaining variables (wrong
procedure). With the correct procedure, we see that there are {\it no
solutions} in which $b$=constant
 \foot{The only exception is $b=\pm 1$. In this case it is easier to see the
solution in the $(\lambda_\pm,\sigma_\pm,b)$ variables, and is given by: (i)
$b=1,\ \lambda_\pm=\lambda_\pm(0),\ \sigma_\pm(\tau)=\sigma_\pm(0)+\tau
\dot\sigma_\pm(0)$ or (ii)
$b=-1,\ \lambda_\pm=\lambda_\pm(0)+\tau\dot\lambda_\pm(0),\
\sigma_\pm(\tau)=\sigma_\pm(0)$. This solution is light-like since it
satisfies $ds^2/d\tau^2=0$}.
 We can also try $u$ or $v$=constant. We find that the only solution of this
type is $u=u_0,\ v=v_0$ and $b(\tau)=\pm cosh[\gamma_\pm (\tau-\tau_\pm)]$ for
$|b|>1$, and $b=cos[\gamma_0(\tau-\tau_0)]$ for $|b|<1$. Here
$u_0,v_0,\gamma_\pm ,\gamma_0, \tau_\pm ,\tau_0$ are integration constants
related to initial conditions or boundary conditions at $b=\pm 1$. One can
picture these geodesics in three dimensions with the help of Fig.1a,b,c and
Fig.4a,b. They are vertical lines parallel to the $b$ axis that may end or
bounce at a singularity at either $b=-1$ or $b=1$ in other regions. Their fate
is determined by the constants $(|v_0|,|u_0|)$ or equivalently $(\rho_\pm ,
\omega_\pm )$. In region I they lie in $1<b<\infty$, in region IIIb they lie
in $-\infty <b<-1$. In region II they may end on the caps of the cylinder or
the caps of the hyperboloids, but if they can avoid those they seem to extend
from $-\infty$ to $+\infty$. Actually they bounce at $b=\pm1$
because $ds^2$ has different signs for $b^2>1$ and $b^2<1$. In region I' they
end at $b=1$, in region III'a,b they end at $b=-1$. In region II' they again
bounce at $b=\pm 1$ even when they are not trapped between the two saddles.
By evaluating the line element $ds^2/d\tau^2=\gamma^2_\pm$ or $-\gamma^2_0$,
we learn that the portion of the geodesic that lies in $-1<b<1$ is time-like
and for $|b|>1$ it is space-like. None of it is light-like.

We see that the direct approach of solving these equations can provide some
information about the space but it is limited. However, as discussed in
section 2, we can find the exact general geodesic solution through the trick
of enlarging the space to the entire $SO(2,2)$ group space ($X_\mu,Y_\mu$)
plus gauge potentials, finding the solution to the differential equations in
the enlarged space and then projecting down to the Lorentz invariants
$Q^a=(v,u,b)$, or equivalently $(\lambda_\pm ,\sigma_\pm , b)$, etc..
Therefore, the first task is to rewrite the solution for the group element in
\sol\ in the form \htt\ and then read off the solution $X_\mu(\tau)$ and
$Y_\mu(\tau)$. This requires evaluating the exponentials in \sol\ in the form
of $4\times 4$ matrices in the vector representation of $SO(2,2)$, which
requires a lot of algebra. This task is a lot easier in the spinor
representation in which it is possible to choose a basis that reduces the
$SO(2,2)$ group element into $2\times 2$ blocks that correspond to the
decomposition $SO(2,2)\rightarrow SL(2,\IR)\times SL(2,\IR)$. In this basis the
exponentials are easy to compute since the generators are represented by
blocks of just $2\times 2$ Pauli matrices. Then by using the group theoretical
correspondence between the spinor and vector representations given below one
can construct the desired solution.

The spinor representation has the added advantage of shedding light on the
global properties of the manifold, including the separation of the manifold to
positive and negative geometrical regions (see below) and the representation
of duality transformations in terms of inversions (see duality section).

Let us first establish a parametrization of the spinor representation and its
correspondance to the vector representation in \htt . We start with the
$4\times 4$ Dirac gamma matrices for three dimensional Minkowski space
$\gamma_\mu=\tau_3\sigma_\mu$, where $\tau_3, \sigma_\mu$ are $2\times 2$
Pauli matrices acting on a direct product space. We choose the basis
$\sigma_\mu=(\sigma_2,i\sigma_1,i\sigma_3)$ which yields
$\{\gamma_\mu,\gamma_\nu\}=2\eta_{\mu\nu}$. From the Dirac matrices one
constructs the 6 generators of $SO(2,2)$ in the $4\times 4$ spinor
representation as follows: $J_\mu={\gamma_\mu\over 2}, \
J_{\mu\nu}={\sigma_{\mu\nu}\over 2}={i\over 4}[\gamma_\mu
,\gamma_\nu]=-\epsilon_{\mu\nu\lambda}{\tau_0\sigma^\lambda\over 2}$, where
$\tau_0$ is the identity Pauli matrix. Thus, the Lorentz subgroup is generated
by $\tau_0\sigma_\mu /2$ and the coset is generated by $\tau_3\sigma_\mu /2$.
By exponentiating these one can construct group elements $h_s,t_s$ in the
spinor representation and put them in the form

\eqn\sht{ h_s={1-i\tau_0 y\over \pm\sqrt{|1+y^2|}}, \qquad t_s={1-i\tau_3x\over
\pm\sqrt{|1+x^2|}}, \qquad y=\sigma_\mu y^\mu ,\quad x=\sigma_\mu x^\mu . }
In this equation we have allowed for the possibility that the Lorentz
invariants $x^2=x^\mu x_\mu,\ y^2=y^\mu y_\mu$ can take any value on the real
line. As required, the determinants of the $4\times 4$ $h,t$ are unity
$det(h_s)=1=det(t_s)$.

To establish a connection to the $4\times 4$ vector representation given in
\htt\ we first define 4 matrices in Dirac space
$V_M=(V_{0'},V_\mu)=(\tau_1\sigma_0 , \tau_2\sigma_\mu)$. They are orthonormal
under the trace, $Tr(V_MV_N)=4\eta_{MN}$, where $\eta_{MN}= diag(1,1,-1,-1)$
is the $SO(2,2)$ metric in the vector representation and can be used to raise
or lower the indices $M=(0',0,1,2)$. Under commutation with the 6 generators
$({\tau_0\sigma_\mu\over 2} , {\tau_3\sigma_\mu\over 2})$ the $V_M$ form the 4
dimensional vector representation of $SO(2,2)$. Therefore, under the action
of any $SO(2,2)$ group element in the spinor representation $g_s$ one finds
that these matrices rotate into each other and form a linear space:
$g_sV_Mg_s^{-1}=(g_v)_M^{\ N} V_N$ . This means that the coefficients
$(g_v)_M^{\ N}$ correspond to the $4\times 4$ vector representation of the
$SO(2,2)$ group element $g_s$ and can be written as

\eqn\vectorrep{ (g_v)_M^{\ N}= {1\over 4} Tr(g_sV_Mg_s^{-1}V^N) . }
This is the construction of the vector representation from the product of two
spinor representations. Applying this map to $h_s,t_s$ given in \sht\
we derive their vector representatives $h_v,t_v$ and compare them to the
expressions $h,t$ in \htt . The inverse matrices needed in this computation
are $h_s^{-1}=h_s(-y)sign(1+y^2)$ and similarly for $t_s^{-1}$. From this
simple algebra we derive the relationship between $(X_\mu , Y_\mu)$ and
$(x_\mu ,y_\mu)$ given in \change .

Any group element may be written in the form $g=ht$ in any representation. For
the spinor representation in our basis this gives a block diagonal form

\eqn\form{ g_s=\pmatrix{g_+ & 0\cr 0 & g_-\cr} \qquad
 g_+={(1-iy)(1-ix)\over \pm\sqrt{|(1+y^2)(1+x^2)|} }, \qquad
 g_-={(1-iy)(1+ix)\over \pm\sqrt{|(1+y^2)(1+x^2)|} }, }
where we see that the determinants of each block
$det(g_\pm)=sign[(1+x^2)(1+y^2)]$ could be $\pm 1$,  while the overall
determinant remains at $det(g)=1$. Therefore, $SO(2,2)$ in the spinor
representation goes {\it globally} beyond $SL(2,\IR)\times SL(2,\IR)$ by
allowing
the determinants of the $2\times 2$ blocks to have the value of $-1$
simultaneosly. As we already saw in the previous sections these signs are
closely tied to the globally separate {\it positive and negative geometrical
regions!} The spinor representation now explains the group theoretical origin
of this fact and tells us that we are going beyond $SL(2,\IR)\times SL(2,\IR)$
when we are in the negative region (double saddle).

We are now in a position to write the spinor representation of the general
geodesic solution for the group element given in \sol\const . The constants of
integration are $x_0^\mu, y_0^\mu, p^\mu, \alpha^\mu$ and they are required to
obey the constraint $(g_0(\tau_3p-\tau_0\alpha)g_0^{-1})_H+\tau_0\alpha=0$,
where $p=i\sigma_\mu p^\mu$ , $\alpha=i\sigma_\mu \alpha^\mu$, and $g_0$ has
the form \form\ with $x_0,y_0$ inserted. The $H$ projection is implemented by
dropping all terms that contain $\tau_3$ after multiplying the factors. Then
the constraint reduces to two $2\times 2$ identical blocks of the form

\eqn\nconst{ (1+x_0^2)^{-1}(i[p,x_0]-\alpha - x_0\alpha x_0) + (1+y_0^2)^{-1}
(i[y_0,\alpha ] + \alpha + y_0\alpha y_0) = 0 . }
In component notation this becomes

\eqn\ealpha{M^{\mu\nu}\a_{\nu}=-r \e^{\mu\nu\lambda}x_{0\nu}p_{\lambda}, \quad
 M^{\m\n}=(1-r)\eta^{\m\n}+y_0^{\m}y_0^{\n}-r x_0^{\m}x_0^{\n}-
\e^{\m\l\n}y_{0\l},
 \quad r={1+y_0^2 \over {1+x_0^2}}\ .  }
Using the notation $(a\times b)^\mu =\epsilon^{\mu\nu\lambda} a_\nu b_\lambda$
one can write the generic solution for $\a^{\m}$ in the form

\eqn\solualpha{\a^{\m}=(x_0\times p)^{\mu} - \big [(x_0\times p)\cdot y_0 \
\big (y_0^\mu + (x_0\times (y_0\times x_0))^\mu\big )-(x_0\times y_0)\cdot
(x_0\times p)\ x_0^\mu \big]/(x_0\times y_0)^2 . }
This expression is valid as long as $(x_0\times y_0)^2\ne 0$. As seen from
\change\inv\scalar\ this means that at $\tau=0$ the particle is not right on
the singularity $v(0)\ne u(0)+2$.
 \foot{ The solution is also easily found when $(x_0\times y_0)^2=0$. For
example when $x_0^\mu$ and $y_0^\mu$ are parallel, $y_0^\mu =\gamma x_0^\mu$
then

\eqn\parallel{ \alpha^\mu=\beta x_0^\mu - {\gamma (1+x_0^2)\over
x_0^2(\gamma^2+x_0^2)} (x_0\times (x_0\times p))^\mu + {\gamma^2-1\over
\gamma^2+x_0^2} (x_0\times p)^\mu , }
where $\beta,\gamma$ are arbitrary constants. }

Next we compute the exponentials in \sol\ in the spinor representation. Since
every factor splits into $2\times 2$ blocks as in \form\ we evaluate each
block separately $g_\pm=e^{\alpha\tau}g_\pm(0) e^{(\pm p-\alpha)\tau}$. Since
$det(e^{\alpha\tau})=det(e^{(\pm p-\alpha)\tau}=1$, an immediate result is

\eqn\sign{ det(g_\pm(\tau))=det(g_\pm(0))= sign((1+x_0^2)(1+y_0^2)) \equiv
\epsilon_0 , }
which shows that geodesics never cross from the positive to the negative
region since the sign $\epsilon_0$ is time independent. Now we compute the
exponentials

\eqn\expon{\eqalign {
 & e^{\alpha\tau}=c_0(\tau)+i\sigma\cdot\alpha\ s_0(\tau),\qquad
 e^{(\pm p-\alpha)\tau}=c_\mp(\tau)-i\sigma\cdot(\alpha\mp p)\
 s_\mp(\tau),\cr
 & c_0(\tau)=cos(\sqrt{\alpha^2}\tau), \qquad
 s_0(\tau)={sin(\sqrt{\alpha^2}\tau)\over \sqrt{\alpha^2} }\cr
 & c_\mp(\tau)=cos(\sqrt{(\alpha\mp p)^2}\tau), \qquad
 s_\mp(\tau)={sin(\sqrt{(\alpha\mp p)^2}\tau)\over \sqrt{(\alpha\mp p)^2} }. }}
The resulting $2\times 2$ matrices $g_\pm (\tau)$ can be rewritten in the form
\form\ in order to read off the solution for $x_\mu(\tau)$ and $y_\mu(\tau)$.
While this can certainly be done, we only need the Lorentz invariants which
can be extracted as follows

\eqn\bls{ \eqalign {
 & det(g_+(\tau)+g_-(\tau))= {4 \epsilon_0 \over 1+x^2(\tau) } = 2\epsilon_0
(b(\tau)+1), \cr
 & Tr(g_+(\tau)+g_-(\tau))= \pm 4 [|(1+x^2(\tau))(1+y^2(\tau)|]^{-{1\over 2}} =
2\lambda_{\epsilon_0}(\tau), \cr
 & Tr(g_-(\tau)-g_+(\tau))= \pm 4x(\tau)\cdot y(\tau)\
[|(1+x^2(\tau))(1+y^2(\tau)|]^{-{1\over 2}} = 2\sigma_{\epsilon_0}(\tau) . }  }
where $\lambda_{\epsilon_0},\sigma_{\epsilon_0}$ are the $\lambda_\pm ,
\sigma_\pm $ global coordinates defined in the previous section, and the value
of $\epsilon_0=\pm$ is determined by the initial conditions as in \sign .
Performing these computations gives the result

\eqn\bbllss{ \eqalign {
 & {\lambda_{\epsilon_0} (\tau)\over \lambda_{\epsilon_0} (0)}=
c_0(\tau)[A_1^+c_+(\tau)+A_1^-c_-(\tau)-A_2^+s_+(\tau)-A_2^-s_-(\tau)] \cr
 &\hskip 2cm
 +s_0(\tau)[A_3^+c_+(\tau)+A_3^-c_-(\tau)+A_4^+s_+(\tau)+A_4^-s_-(\tau)]  \cr
 & {\sigma_{\epsilon_0} (\tau)\over \pm \lambda_{\epsilon_0} (0)}=
c_0(\tau)[A_1^+c_+(\tau)-A_1^-c_-(\tau)-A_2^+s_+(\tau)+A_2^-s_-(\tau)] \cr
 &\hskip 2cm
  +s_0(\tau)[A_3^+c_+(\tau)-A_3^-c_-(\tau)+A_4^+s_+(\tau)-A_4^-s_-(\tau)]  \cr
 & {b(\tau)\over b(0)}= c_+(\tau)c_-(\tau)+A_5s_+(\tau)s_-
(\tau)+A_6^+s_+(\tau)c_-(\tau)+A_6^-s_-(\tau)c_+(\tau) . } }
where the various constants are determined by the initial parameters as follows

\eqn\aaaa{\eqalign {
 & \lambda_{\epsilon_0}(0)={\pm 2\over \sqrt{|(1+x_0^2)(1+y_0^2)|} }, \qquad
  b(0)={1-x_0^2\over 1+x_0^2} , \qquad A_1^\pm={1\over 2}(1\pm y_0\cdot x_0)\cr
 & A_2^\pm={1\over 2}(\alpha\pm p)\cdot (y_0\mp x_0 \mp y_0\times x_0), \qquad
  A_3^\pm={1\over 2}\alpha \cdot (y_0\mp x_0 \mp y_0\times x_0), \cr
 & A_4^\pm={1\over 2}[(1\pm y_0\cdot x_0)(\alpha^2\pm\alpha\cdot p)\mp
(\alpha\times p)\cdot (y_0\mp x_0 \mp y_0\times x_0)], \cr
 & A_5=\alpha^2-p^2 - {4\alpha\cdot (x_0\times p)\over 1-x_0^2}, \qquad
   A_6^\pm= {2 x_0\cdot (p\pm \alpha)\over 1-x_0^2} . } }
Using \newco\ we have explicitly checked that the geodesic equations \geod\
are indeed satisfied by the above general solution.

Depending on initial positions and velocities, the arguments of the functions
$c_0,c_\pm , s_0,s_\pm $ may turn out to be real or imaginary, as determined
by the $signs(\alpha^2, (\alpha + p)^2, (\alpha -p)^2)=(\pm,\pm,\pm)$.
Accordingly, the solutions may contain oscillating trigonometric functions or
their hyperbolic counterparts. The far past and far future position of
particles crucially depend on these signs. In the purely trigonometric case,
$signs=(+,+,+)$, the particle oscillates in the vicinity of the singularity
surface and cannot escape from its gravitational pull.  The curvature scalar
\scalar\ is not zero and the particle never reaches the asymptotically flat
region. It turns out that this kind of initial condition is possible only
in region II, and not in the others. In this purely oscillating solution
the particle wobbles around the central blob of Fig.3a and nearby
regions of the trousers. When one or two of the signs are negative the
particle can get far away from the singularity surface temporarily but
periodically returns to parts of it. By computing the
$\tau\rightarrow\pm\infty$ asymptotic behavior of the trajectories one finds
that the scalar curvature does not vanish, and therefore the particle does not
reach the asymptotically flat region. Finally, when all signs are negative,
$signs=(-,-,-)$, only hyperbolic functions occur in the solution and the
particle is found only in the flat regions in the far past and far future of
its lifetime. This result is obtained by computing the asymptotic behavior of
\bbllss\ for $\tau\rightarrow\pm\infty$

\eqn\asympt{\eqalign {
 & \lambda_{\epsilon_0}(\tau)\rightarrow
 A^+\ e^{(\sqrt{|\alpha^2|}+\sqrt{|(\alpha+ p)^2|})|\tau|}
 + A^- e^{(\sqrt{|\alpha^2|}+\sqrt{|(\alpha- p)^2|})|\tau|} \cr
 & \sigma_{\epsilon_0}(\tau)\rightarrow
 \pm A^+\ e^{(\sqrt{|\alpha^2|}+\sqrt{|(\alpha+ p)^2|})|\tau|}
 \mp A^- e^{(\sqrt{|\alpha^2|}+\sqrt{|(\alpha- p)^2|})|\tau|} \cr
 & b(\tau)\rightarrow B\ e^{\sqrt{|(\alpha
+p)^2|}+\sqrt{|(\alpha - p)^2|})|\tau|}  } }
where

\eqn\asscoeff{\eqalign {
 & A^\pm(sign(\tau))={\lambda_{\epsilon_0}(0)
\over 4}\big [A_1^\pm+{A_4^\pm\over
\sqrt{\alpha^2(\alpha\pm p)^2} } + sign(\tau)({A_3^\pm\over
\sqrt{|\alpha^2|}}- {A_2^\pm\over \sqrt{|(\alpha\pm p)^2|} }) \big ] \cr
 & B(sign(\tau))={b(0)\over 4} \big [ 1+{A_5\over \sqrt{(\alpha+p)^2(\alpha-
p)^2}}
+sign(\tau)({A_6^+\over \sqrt{|(\alpha+p)^2|}}+{A_6^-\over \sqrt{|(\alpha-
p)^2|}})\big ]. }     }
Comparing to the discussion of the asymptotically flat region in
\flat\ we see that indeed, in the purely hyperbolic case, the particle {\it
must} escape the gravitational pull of the singularity at large past and
future times. This type of behavior is possible in all regions
I,II,III,I',II',III' through a choice of the initial values at $\tau=0$.

   From this analysis we arrive at the following important conclusion: A
particle which is found in the flat region must have hyperbolic initial
conditions $signs=(-,-,-)$ since otherwise it could not be there. Thus, if a
particle starts out in the flat region, and travels toward the singularity, it
must return to another part of the flat region after some time. We may now ask
where does such a particle go during its journey? For this we need to
discuss the initial conditions as follows.

Substituting $i\tau_3\sigma_\mu p^\mu$ for the matrix in \dsds\ one finds a
fixed value for the line element (or Lagrangian) associated with the solution
given above

\eqn\dsdsds{ {ds^2\over d\tau^2}=-2p^\mu p_\mu . }
This allows one to easily control the signature of the geodesic by choosing
time-like, space-like or light-like momenta $p^\mu$ as an initial condition.
The remaining initial conditions $(x_0^\mu ,y_0^\mu)$ can also be chosen
according to the region one wishes to explore. Table.1 is useful for this
purpose. Once one picks one of the regions I,II,III,I',II',III' one can
position oneself in it by first choosing $(v(0),u(0),b(0))$, which is
equivalent to a choice of parameters and $\epsilon, \epsilon'$ from Table.1.
This determines the vectors $X^\mu(0),Y^\mu(0)$ from which we deduce
$x_{0\mu}=-X^\mu(0)b(0)/(1+b(0))$ and $y_0^\mu=Y^\mu(0)/v(0)$ in a particular
Lorentz frame. We can then compute all of the Lorentz invariant constants in
\aaaa\ that determine the geodesics \bbllss .

We have written Mathematica and Lotus programs with the above inputs and
plotted the geodesics by taking various initial conditions at $\tau=0$. We
have then examined the location of the particle at both negative and positive
values of $\tau$. These numerical plots reveal very interesting behavior
in the vicinity of the singularity. In the purely hyperbolic case, at large
negative proper times (far past) the particle is far away from the singularity
at large values of $(\lambda_{\epsilon_0},\sigma_{\epsilon_0},b)$. In a finite
amount of proper time the particle approaches the singularity {\it
tangentially} and bounces off from it. Depending on the initial conditions
chosen, this may happen several times at various parts of the singularity
(i.e. at the legs or body of the pants in the positive region, or at the
saddles in the negative region). After a finite amount of proper time, the
particle leaves the singularity region and returns to large values of
$(\lambda_{\epsilon_0}, \sigma_{\epsilon_0}, b)$ at large positive proper
times (far future).

Such a hyperbolic trajectory is quite interesting, especially when contrasted
to the trajectory of a particle that falls into a black hole. In the case of a
black hole singularity a particle that falls in never comes back and also
cannot send any signals once it passes the horizon. However, in the present
case, a particle can start out far away from the singularity in the flat
region, fall in, gather information from the neighborhood of the singularity,
and come back to another part of
the flat region after a finite amount of proper time.
Therefore, the notion of a ``horizon", if any, is quite different than the
case of a black hole singularity.

If the initial conditions are not purely hyperbolic the numerical plots
confirm that the particle is either partially or completely trapped by the
singularity as described above. The particle trajectory bounces off various
parts of the singularity, and never sticks to it, unlike a black hole. An
intuitive physical reason for not sticking to a special point of the
singularity is the gravitational attraction of all the other points that form
the singularity. Namely, the singularity at any given instant of time is a
string, not a point. Therefore, the rest of the string exerts a gravitational
force on the particle, thereby not allowing the particle to come to rest at
any special point on the string.

An interesting question is whether there might exist closed time-like curves
in any of the regions of our global space? To search for these one must allow
for the possibility that an external force (such as the firing of a rocket on
the spaceship) might change the course of the trajectory at some value of the
proper time $\tau$. This is expressed as a change in $p_\mu\rightarrow p'_\mu$
at some value $\tau=\tau_1$. The evolution of the particle trajectory can be
computed according to \sol\const\ during $\tau_0<\tau<\tau_1$, while for
$\tau_1<\tau$ one again uses the same rules, but with initial conditions
$g(\tau_1),p'$. The external forces may act more than once.
The question of closed time-like trajectories boils down to
whether, for positive $p^2,p^{'2},\cdots $, there exists a $\tau_2$ such that
$g(\tau_2)=\Lambda g_0\Lambda_{-1}$, where $\Lambda$ is a gauge
transformation? We have been unable to answer this question conclusively.
However, after staring at many numerical plots, we conjecture that in regions
I,III,I',III' closed time-like curves do not seem possible. We are less sure
about regions II,II'. However, if closed time-like curves are at all possible
in one of these regions, it could happen only to an observer that is
travelling in the vicinity of the singularity, not while he is in the flat
region. Such a closed time-like curve would not allow this observer to ``kill"
his mother just before he is born unless his mother was also in the vicinity
of the singularity at the time of the birth as well as at the time of the
murder. Then this, if it occurs, seems to have no consequence on observers
that are located in the flat region. Therefore, it does not seem possible that
causality in the flat regions can be violated. This is probably sufficient not
to get into trouble with causality as we know it in flat regions.

As noted earlier in the text, the metric and dilaton discussed in this paper
are conformally exact only for  the type-II superstring (and any type of point
particle theory). For the purely bosonic and heterotic strings the conformally
exact metric and dilaton have also been computed for the 3d and 4d models
based on $SO(2,2)$ and $SO(3,2)$ \BAS . The exact versions have singularity
surfaces and regions whose properties differ in interesting ways from those
discussed in this paper. These results will be given in a separate
publication. We emphasize that the geodesic analysis discussed in this section
is valid in the purely classical limit in which the dilaton is neglected.

\newsec{ Comments on possible physical applications }

We have shown that a global analysis can readily be given for all geometries
that arise from gauged WZW models. These are interesting for both particle and
string theories and in either case the small-large duality property
is a novel feature worth of further study. We emphasize that, contrary to
common belief, duality is not only a string property, since the particle
action \action\ shares this feature. In string theory with one time
coordinate the coset models listed in the introduction
are of special interest since they are the only known curved space-time
cases for which there is a chance, at least in principle, of solving the
conformal field theory through current algebra methods. In particle theories
these models are also very special since the quantum spectrum can be
computed exactly through group theoretical representation theory for
non-compact groups. This presents a rare opportunity in General Relativity
investigations.

The specific 3d model which we have investigated in more detail has a
cosmological interpretation. This was already apparent from the remarks
following eq.\flat . If we imagine that the ``Big Bang" was not a point
singularity, but a string singularity, and that cosmological time begins near
the central blob of Fig.3a, then matter and energy that was created initially
will have a future that is determined by the initial location and initial
velocity. Let us first consider region I or III, assuming that the ``Big Bang"
corresponds to the pinch at one of the trouser's legs. Inside a trouser's leg
all particle trajectories are purely hyperbolic and therefore, all the matter
and energy created by the ``Big Bang"
will eventually travel to the asymptotically
flat region. Thus, this singularity may be considered a cousin of  a ``white
hole". The picture is somewhat different for region II. We have seen that in
this region all signs are possible for $signs=(\pm,\pm,\pm)$. Therefore,
depending on initial conditions, some of the matter and energy will remain
trapped (dark matter ?)  while some other part will escape to the
asymptotically flat region. It would be interesting to pursue such models for
cosmological applications.

In our discussion of $SO(2,2)/SO(2,1)$ we assumed that we were dealing with
a 3d theory. However, we can adjoin a factor of $U(1)$ or $\IR$ as a fourth
flat dimension, and our entire discussion would then apply to the 3d subspace
of a four dimensional theory. Since the $U(1)$ or $\IR$ factor absorbs one
unit of the Virasoro central charge, $c=1$, we must choose $k=5/2$ for the
$SO(2,2)_{-5/2}$ current algebra to produce the balance $c=25$. For a
similar supersymmetric theory in 4 dimensions the flat dimension produces
$c=3/2$, therefore $k=3$ so that with $SO(2,2)_{-3}$ the total central charge
is $c=15$. The fact that $k=3$ is an integer in this last case may be
significant from the point of view of global anomalies.

It was shown in \BASF\ that a heterotic string model can be constructed
directly in 4 curved space-time dimensions. The super coset in this case is
$SO(3,2)/SO(3,1)$ at level $k=5$. The perturbative metric and dilaton were
given in one of the patches. Clearly, it would be desirable to work out the
global analysis of this geometry. One of the interesting features of that model
was that it admitted $SU(3)\times SU(2)\times U(1)$ as the flavor symmetry
group at level 1. Therefore, one expects a certain number of families of quarks
and leptons to emerge as $SU(3)$ triplets and $SU(2)$ doublets.

It is a general hope that a vacuum configuration of the heterotic string
correctly describes the low energy spectrum of quark and lepton families.
We think that, this notion is more attractive and more believable when the
vacuum configuration of the string describes a
time dependent {\it cosmological} curved space-time.
Then one can imagine that the quarks and leptons were produced at the initial
``Big Bang", in the presence of strong gravitational fields (more realistic
than flat space), and then travelled along geodesics
to flat regions of space where
they are observed today. Therefore, 4d curved space-time models of the type
described above are very interesting to study these notions. As it was made
clear in the introduction, there are a small number of
4 dimensional conformally exact
heterotic string models based on non-compact cosets. Taken, with a
cosmological interpretation as above, such models have the potential to
describe quantum mechanically the fundamental matter at the earliest times. We
are in the process of studying the quantum theory of these models and hope to
report on these issues in future publications.


\newsec{ Acknowlegements}

We would like to thank heartily N. Warner for his general support and
especially for making his NeXt computer and his office available while we
performed graphical and numerical computations. We would also like to thank
our colleagues K. Pilch and A. Cooper for discussions and insights.


\vfill\eject
\centerline {\bf Appendix }
\bigskip

The reader who is familiar with the 2-dimensional $SO(2,1)/SO(1,1)$ manifold
that describes a black hole, may find it useful to see it written in our
notation and then compare it to the 3-dimensional spaces described above. The
2d theory is reached from the group element in \htt\ by specializing $X_\mu$
to a 2d vector and taking
$h_{\mu\nu}=\epsilon'cosh(\sqrt{u})+\epsilon_{\mu\nu}sinh(\sqrt{\pm u})$. As
our Lorentz invariant global coordinates we take $Q^a=(u,b)$ where $b$ was
defined in \htt . The global metric and dilaton are then given by

\eqn\metricc{ ds^2={db^2\over 4(b^2-1)} - {b-1\over b+1}{du^2\over 4u},\qquad
\Phi=ln(b^2-1)+\Phi_0. }
The global space for $SO(2,1)/SO(1,1)$ is $(u>0,\ -\infty<b<\infty)$, as shown
in Fig.2 . For this case we may rewrite $t=\pm\sqrt{u},\ du^2/4u=dt^2$ where
$-\infty<t<\infty$ is a non-compact coordinate which takes the role of time in
the regions marked $I(-+), IIa(-+)$ and the role of space in the region marked
$IIb(+-)$. The regions IIa and IIb are geodesically connected, with IIa outside
the horizon and IIb within the horizon, with the black hole at $b=-1$. The
region marked $I(-+)$ is geodesically isolated from IIab and contains a naked
singularity at $b=-1$. All this is easy to see by transforming to the Kruskal
coordinates that are also global coordinates, $b=1-2v_+v_-,\ v^2_{\pm}={1\over
2} |b-1| e^{\pm t}$, and for which the metric takes the standard form
$ds^2=dv_+dv_-/(v_+v_--1)$ \WIT . Note that in the $(u,b)$ parametrization the
various regions of the $(v_+,v_-)$ manifold correspond to 2 semi-infinite
sheets that extend toward positive $u$ (since $t$ has both signs), sewn
together at $u=0$ along the $b$ axis, and cut at $b=-1$ along the singularity.

Fig.2 also shows the region  $(u<0, \ -\infty<b<\infty )$ which corresponds
to the analytic continuation of the coset $SO(2,1)/SO(1,1)$ to $SO(2,1)/SO(2)$
or $SO(3)/SO(2)$. In these regions we may write $h_{\mu\nu}=cos\theta +
\epsilon_{\mu\nu} sin\theta$ where $\theta$ is a compact coordinate $0<\theta
<2\pi$ (note that Minkowski-Euclidean analytic continuation also requires
$\epsilon_{01}\rightarrow i\epsilon_{21}$). The metric then describes a cigar
a cymbal and a trumpet in the indicated regions \WIT
 \ref\euclidean{S. Elitzur, A. Forge and E. Rabinovici,
 Nucl. Phys. {\bf B359} (1991) 581. }.
In the cigar region $b=cosh(2r),\ r>0$,  and $ds^2=dr^2+tanh^2r\ d\theta^2$.
In the trumpet region $b=-cosh(2r'), \ r'>0$ and $ds^2=(dr')^2+coth^2r'\
d\theta^2$. In the cymbal region $b=cos(2r''),\ 0<r''<\pi /2$ and $ds^2=-
(dr'')^2-tan^2r''\ d\theta^2$ (in this region we need to change $k\rightarrow
-k$ to get $(++)$ signature). The tip of the cigar touches the zenith of the
cymbal at $b=1$, while the trumpet and cymbal touch their (infinite) skirts at
$b=-1$ . In the $(u,b)$ parametrization all of these shapes have been deformed
to double sheeted strips  that lie parallel to the $b$ axis in the $u$-range
$-(2\pi)^2<u<0$ and sewn together at $u=0,-(2\pi)^2$ (periodic in
$0<\theta<2\pi$). Furthermore the strips are cut at $b=-1$ while sewn at
$b=1$. Further out regions toward negative $u$ repeat periodically the same
(cigar,cymbal,trumpet) ``music". The Minkowski-Euclidean analytic continuation
is $SO(1,2)/SO(1,1)\rightarrow SO(2,1)/SO(2)$ or $SO(2,1)/SO(1,1)\rightarrow
SO(3)/SO(2)$ which corresponds to analytic continuation from positive to
negative $u$ in Fig.2 .

Let us briefly review the duality properties of the 2d manifold in our
notation and point out a new feature. Switching the sign $\epsilon$ is
equivalent to $(u',b')=(u,-b)$ for the 2d global coordinates. Fig.2 then shows
that duality flips (I,IIa) and (cigar,trumpet), while the regions (IIb) and
(cymbal) are self dual. In this process the singularity and the horizon also
get interchanged. The group theoretical meaning of this transformation is
understood by examining the group element in \htt\ (specialized to 2d): under
duality the group element takes a leap in group space. Evidently, this leap
does not change the theory, it only rearanges the regions. To better
understand what is going on it is useful to make a change of coordinates
$X_\mu =2x_\mu/(x^2-1)$ which allows one to write $b=\epsilon (1+X^2)^{-
{1\over 2}} = (1-x^2)/(1+x^2)$. The invariant $x^2$ is anywhere on the real
line $-\infty<x^2<\infty$. The duality transformation $b'=-b$ is now generated
by $x'_\mu=-x_\mu/x^2$ which corresponds to an inversion in $x_\mu$ space.
Under this inversion $X_\mu$ remains invariant but the group element makes
just the required leap. This new version of duality is remarkably similar  to
the one encountered in tori ($R\rightarrow 1/R$) or mirror manifolds (see
also footnote 1).

\vfill\eject
\centerline {\bf Figure Captions}
\bigskip

Fig.1-- Global spaces for $SO(2,2)/SO(2,1)$, $SO(3,1)/SO(2,1)$, $SO(3,1)/SO(3)$
and $SO(4)/SO(3)$, and dual patches.

Fig.2-- Global spaces for $SO(2,1)/SO(1,1)$, $SO(2,1)/SO(2)$ and
$SO(3)/SO(2)$, and dual patches.

Fig.3a-- Pinched double trousers singularity in the positive sector, with
$(\lambda_+,\sigma_+,b$ coordinates.

Fig.3b-- Double saddle singularity in the negative sector, with $(\lambda_-
,\sigma_-,b)$ coordinates.

Fig.4a-- Pinched double trousers singularity in the positive sector, with
$(\rho_+,\omega_+,b)$ coordinates.

Fig.4b-- Double saddle singularity in the negative sector, with $(\rho_-
,\omega_-,b)$ coordinates.

\listrefs
\end